\documentclass[aps,pro,superscriptaddress,floatfix]{revtex4}
\usepackage{amsfonts,color}
\usepackage{amsmath}
\usepackage{amssymb}
\usepackage{graphicx,dsfont}
\usepackage{multirow}
\usepackage{ulem}
\usepackage{datetime2}

\usepackage[
colorlinks,
linkcolor=blue,
urlcolor=blue,
citecolor=blue,
plainpages=false,
pdfpagelabels,
]{hyperref}

\begin{document}

\title{Trigonometric $SU(N)$ Richardson-Gaudin models and dissipative
multi-level atomic systems}
\author{Sergio Lerma-Hern\'andez}
\affiliation{Facultad de F\'{\i}sica, Universidad Veracruzana, Circuito Aguirre Beltr\'an
s/n, Xalapa, Veracruz 91000, Mexico}
\author{Alvaro Rubio-Garc\'ia}
\affiliation{Instituto de Estructura de la Materia, CSIC, Serrano 123, 28006 Madrid, Spain}
\author{Jorge Dukelsky}
\affiliation{Instituto de Estructura de la Materia, CSIC, Serrano 123, 28006 Madrid, Spain}

\begin{abstract}
We derive the exact solution of a system of $N$-level atoms in contact with
a Markovian reservoir. The resulting Liouvillian expressed in a vectorized
basis is mapped to an $SU(N)$ trigonometric Richardson-Gaudin model whose
exact solution for the complete set of eigenmodes is given by a set of
non-linear coupled equations. For $N=2$ ($SU(2)$) we recover the exact
solution of Phys. Rev. Lett. 122, 010401 (2019). We then study the $SU(3)$
case for three-level atom systems and discuss the properties of the steady
state and dissipative gaps for finite systems as well as for the thermodynamic limit.
\end{abstract}

\maketitle

\section{Introduction}

Richardson-Gaudin (RG) integrable models can be traced back to the original
works of Richardson for the exact solution of the BCS model \cite%
{Richardson1963} and Gaudin for the derivation of the integrable quantum
magnet \cite{Gaudin}. Both exact solutions were later recovered in the
context of ultrasmall superconducting grains \cite{Sierra2000} and further
combined to give birth to families of exactly solvable models based on the
rank 1 algebras $SU(2)$ for fermions and $SU(1,1)$ for bosons \cite{Duke2001,
	Amico2001} (for a review see \cite{Colloquium, NuclPhysB}). With a few exceptions these families of exactly solvable models were
extensively applied to closed mesoscopic systems with hermitian
Hamiltonians in different areas of many-body quantum physics, like nuclear
physics, condensed matter, cold atoms, quantum optics and quantum chemistry.
Extensions of the RG integrable models to higher rank algebras include $SO(5)$
\cite{Links_2002, O5} and $SU(3)$ \cite{SU3} (rank 2), $SO(6)\equiv$ $SU(4)$ \cite{Links, O6}
(rank 3) and $SO(8$) \cite{O8}(rank 4). Despite the fact that most of
the applications focus on the study of hermitian Hamiltonians, RG models
can incorporate non-hermiticity by either using complex coefficients for
the linear combination of the integrals of motion, or even integrals of
motion with complex internal parameters. Both cases produce exactly solvable
non-hermitian operators that have been exploited to study non-hermitian
pairing Hamiltonians \cite{Kaneko, Plo1, Plo2}. However, non-hermiticity has a much broader area of application in many-body quantum physics. In particular, it is a
distinctive feature of the dynamics of dissipative quantum systems. The time-dependent density matrix of quantum systems weakly coupled to a Markovian environment is governed by the Lindblad master equation. This master equation defines a non-hermitian Liouvillian superoperator \cite{Book} that, as we will see below, acts on the space of density matrices.
There are very few examples of exactly solvable Liouvillians, mostly based on a tensor network ansatz \cite{Prosen2011, Karevski2013} or on the Bethe
ansatz \cite{PhysRevX.7.041015, Medvedyeva, Ziolkowska, Shibata, Ueda, Buca}. Similarly, two recent
contributions showed that Liouvillians describing either a chain of 1/2
spins \cite{Lamacraft} or a single collective spin \cite{Rib-Pro} coupled to
a Markovian environment are exactly solvable $SU(2)$
RG superoperators. The aim of this paper is to extend the realm of
exactly solvable RG models to dissipative systems of $N$-level
atoms using the $SU(N)$ trigonometric RG models. These integrable models, that have been formally introduced in references \cite{Jurco89, Ushveridze94, SU3, ASOREY2002593}, did not find find up to now a physical application.

We will start in Section II by setting up the Liouvillian superoperator of
$N$-level atom systems. Section III introduces the $SU(N)$ RG trigonometric
models. In Section IV we show that the Liouvillian superoperator of $N$-level atom systems can  be
expressed as a linear combination of the $SU(N)$ RG integrals of motion. We
derive the RG equations and obtain the eigenvalues
of the integrals of motion and the Liouvillian for each independent
solution. In Section V we derive the exact solution for two-level systems,
which is equivalent to the solution of Ref. \cite{Rib-Pro} for a collective spin. Section VI is devoted to a detailed study of the exact solution for dissipative three-level systems. In Section VII we generalize the Schwinger boson mean-field theory for multi-level dissipative atom systems. With this technique we study the thermodynamic limit (TL) of the $SU(3)$ RG model and compare the results with large scale exact numerical solutions.

\section{Dissipative $N$-level atom systems}

Let us start with a non-interacting system of $L$ $N$-level atoms
characterized by the $U(N)$ generators
\begin{equation}
K_{\alpha \beta }=
\sum_{i=1}^{L}|\alpha\rangle_{_{\!\mbox{\footnotesize $i$}}}{}_{_{\mbox{\footnotesize $i$}}}\!\langle \beta |
 \label{K}
\end{equation}%
where 
 $i$ labels each of the $L$ atoms and Greek letters refer to the
atom levels ($\alpha =1,\cdots ,N$). The set of $N^{2}$ operators ($\ref{K}$%
) close the $U(N)$ commutator algebra
\begin{equation}
\left[ K_{\alpha \beta },K_{\gamma \delta }\right] =\delta _{\beta \gamma
}K_{\alpha \delta }-\delta _{\alpha \delta }K_{\gamma \beta }.
\end{equation}

The linear Casimir operator $C^1\equiv\sum_{\alpha =1}^{N}{K}_{\alpha \alpha
}=L$ is a conserved quantity that reduces the number of independent
operators to $N^2-1$, satisfying the $SU(N)$ commutator algebra.

The Hamiltonian of the non-interacting system
\begin{equation}
H=\sum_{\alpha }\varepsilon _{\alpha }K_{\alpha \alpha }
\end{equation}%
acts on an irreducible representation (\textit{irrep}) of $SU(N)$. The highest weight (HW)
state $|\Lambda_1\rangle=|L,0,\cdots,0\rangle$, with all atoms in their lowest state $\alpha =1$,  satisfies $
K_{\alpha\beta}|\Lambda_1\rangle=0$ for all $\alpha<\beta$. Within this \textit{irrep} the quadratic
Casimir operator of $SU(N)$ reduces to

\begin{equation}
{C}^{2} = \sum_{\alpha,\beta=1}^{N}{K}_{\alpha \beta }{K}_{\beta \alpha}
= L^{2}+\left( N-1\right) L.
\end{equation}
Both Casimirs $C^1$ and $C^2$ are conserved quantities, since they
trivially commute with the Hamiltonian.

We assume that the atomic system is weakly coupled to an environment
fulfilling Markovian conditions. Under these conditions the time evolution
of the density matrix $\rho$ of the system is given by the Lindblad master
equation \cite{Book}
\begin{align}
\frac{d\rho }{dt} =\mathcal{L}\rho =-\mathrm{i}\left[ H,\rho \right] +\sum_{\alpha
	,\beta }\left[ W_{\alpha \beta }\,\rho\, W_{\alpha \beta }^{\dagger }-\frac{1}{2}%
W_{\alpha \beta }^{\dagger }W_{\alpha \beta }\,\rho -\frac{1}{2}\rho\, W_{\alpha
	\beta }^{\dagger }W_{\alpha \beta }\right],
\label{Lindblad}
\end{align}%
where we assume that the interaction of the atoms with the environment is
described by generic collective jumps $W_{\alpha \beta }=\sqrt{x_{\alpha
		\beta }}K_{\alpha \beta }$. The master equation also defines the Liouvillian
superoperator, an operator that acts on the space of density matrices with dimension $\mathcal{N}^2$, where $\mathcal{N}$ is the dimension of the Hamiltonian Hilbert space.
The formal solution of equation (\ref{Lindblad}) for time-independent Liouvillians is
\begin{equation}
\rho \left( t\right) =e^{\mathcal{L}t}\rho \left( 0\right).
\end{equation}

Since the Liouvillian superoperator is, in general, non-hermitian and the dissipative terms are semidefinite-negative, it can be shown that its eigenvalues lie in the real non-positive complex plane with at least one 0 eigenvalue. The density matrix, or set of density matrices in case of degeneracies, corresponding to the 0 eigenvalue defines the steady state (SS), to which the system decays in the long time limit. The rest of the eigenvalues are either real negative or complex conjugate pairs with negative real part (decay modes). The typical decay time is determined by the non-zero eigenvalue with the real part closest to 0 and the absolute value of the real part defines the dissipative gap. The inverse of the gap determines the slowest relaxation dynamics in the long-time limit \cite{Minganti2018spectral}.

In what follows it will be convenient to work with the vector representation
of the Lindblad equation \cite{Yoshioka2019, Zambrini2020}. To do so, we double the Hilbert space $\mathcal{H}$ of dimension $\mathcal{N}$ by mapping
the density matrix $\widehat{\rho }$ into the space $\mathcal{H}\otimes \mathcal{H}$ of dimension $\mathcal{N}%
^{2}$
\begin{equation}
\widehat{\rho }=\sum_{\alpha \beta }\rho _{\alpha \beta }\left\vert \alpha
\right\rangle \left\langle \beta \right\vert \rightarrow |\rho
\rangle\rangle= \sum_{\alpha \beta }\rho _{\alpha \beta }|\alpha \beta
\rangle\rangle.
\end{equation}
In a similar manner, we map operators acting to the left or right of the density matrix to superoperators acting on the space of vectorized density matrices as
\begin{equation}
O\,\widehat{\rho }\rightarrow O\otimes I\,|\rho \rangle\rangle,\quad\widehat{\rho }\,O\rightarrow I\otimes O^{T}|\rho \rangle\rangle.
\end{equation}
Within this procedure we double all operators, thus defining a new set of
$SU(N)$ generators $\overline{J}_{\alpha \beta} = I\otimes K_{\alpha \beta}$ in the dual space. The vector form of the Liouvillian superoperator in the $su(N)\otimes su(N)$
space is
\begin{align}
\mathcal{L} =-i\sum_{\alpha =1}^{N}\varepsilon _{\alpha }\left( K_{\alpha
	\alpha }-\overline{J}_{\alpha \alpha }\right) +\sum_{\alpha ,\beta
	=1}^{N}x_{\alpha \beta }\left[ K_{\alpha \beta }\overline{J}_{\alpha \beta }-%
\frac{1}{2}\left( K_{\beta \alpha }K_{\alpha \beta }+\overline{J}_{\beta
	\alpha }\overline{J}_{\alpha \beta }\right) \right].  \label{Liuv}
\end{align}

In order to facilitate the mapping of the Liouvillian to the RG models it is convenient to perform a canonical transformation on the
generators in the dual space that preserves the $su(N)$ algebra $J_{\alpha
	\beta }=-\overline{J}_{\beta \alpha }$, yielding
\begin{eqnarray}
\mathcal{L} =-i\sum_{\alpha =1}^{N}\varepsilon _{\alpha }\left( K_{\alpha
	\alpha }+J_{\alpha \alpha }\right) -\sum_{\alpha ,\beta =1}^{N}x_{\alpha
	\beta }\left[ K_{\alpha \beta }J_{\beta\alpha}+\frac{1}{2}\left( K_{\beta
	\alpha }K_{\alpha \beta }+J_{\beta \alpha }J_{\alpha \beta }\right) \right] .
\label{LiuvT}
\end{eqnarray}
It is easy to check that $\mathcal{L}$ commutes with the\ $N$ operators
\begin{equation}
S_{\alpha }=K_{\alpha \alpha }-\overline{J}_{\alpha \alpha }=K_{\alpha
	\alpha }+J_{\alpha \alpha },  \label{Cs}
\end{equation}
whose eigenvalues, $s_{\alpha }$, define the set of quantum numbers labeling
the invariant subspaces of $\mathcal{L}$, and thus the different sectors in
which the eigenvalues of $\mathcal{L}$ are grouped. These quantum numbers are integers satisfying  the constraints $-L \leq s_\alpha\leq L$ and $\sum_{\alpha}^N s_\alpha=0$.

The Liouvillian ($\ref{LiuvT}$) is in general non-integrable. However, we
will see that for the following restricted set of jump operators the
resulting Liouvillian belongs to the trigonometric family of $SU(N)$
RG models
\begin{equation}
x_{\alpha \beta }=\left\{
\begin{array}{cc}
\Gamma _{0} & \text{for \ }\alpha =\beta \\
\Gamma \left( 1-p\right) & \text{for \ }\alpha >\beta \\
\Gamma \left( 1+p\right) & \text{for \ }\alpha <\beta%
\end{array}%
\right. ,  \label{restriction}
\end{equation}
where $\Gamma$ and $\Gamma_0$ determine the decoherence rate and the
polarization factor $p$ introduces an imbalance between the rising and lowering transitions
induced in the atoms by the environment. Inserting these restrictions into the
Liouvillian (\ref{LiuvT}) and making use of the quadratic Casimir operator
and the commutator algebra, we can express it as the sum of two terms, $%
\mathcal{L}=\mathcal{L}_{C}+\mathcal{L}_{RG}$, where
\begin{equation}
\mathcal{L}_{C}=-\mathrm{i}\sum_{\alpha }\varepsilon _{\alpha }\left( K_{\alpha
	\alpha }+J_{\alpha \alpha }\right) -\Gamma C^{2}+\frac{\Gamma -\Gamma _{0}}{2%
}\sum_{\alpha }\left( K_{\alpha \alpha }+J_{\alpha \alpha }\right) ^{2}
\label{L_C}
\end{equation}%
and
\begin{align}
\mathcal{L}_{RG} = \frac{\Gamma p}{2}\sum_{\alpha }\left( N+1-2\alpha
\right) \left( K_{\alpha \alpha }-J_{\alpha \alpha }\right)-\Gamma
\sum_{\alpha }K_{\alpha \alpha }J_{\alpha \alpha }-\Gamma \left( 1-p\right)
\sum_{\alpha >\beta }K_{\alpha \beta }J_{\beta \alpha } -\Gamma \left(
1+p\right) \sum_{\alpha <\beta }K_{\alpha \beta }J_{\beta \alpha } .
\label{L_RG}
\end{align}%
The term $\mathcal{L}_{C}$ is a conserved quantity depending on the
quadratic Casimir $C^{2}$ and the $N$ operators $S_\alpha$~\eqref{Cs}.
In the next Sections we will introduce the trigonometric $SU(N)$
RG models and show that $\mathcal{L}_{RG}$ can be expressed as
a linear combination of the RG integrals of motion.

\section{Trigonometric $SU\left( N\right) $ Richardson-Gaudin models}

Here we follow the derivation of the $SU\left(3\right)$ RG model in \cite%
{SU3}, and generalize it to $su\left( N\right) $ algebras. The integrals of
motion of the $XXZ$ trigonometric model for $M$ copies of $su(N)$ are
\begin{equation}
R_{m} =\sum_{\alpha =1}^{N}\chi _{\alpha }K_{\alpha \alpha
	m}+\sum_{m^{\prime }\left( \neq m\right) =1}^{M}\left[ Z_{m^{\prime
	}m}\sum_{\alpha =1}^{N}K_{\alpha \alpha m}K_{\alpha \alpha m^{\prime }}+
\sum_{\beta >\alpha }\left(X_{m^{\prime }m}K_{\alpha \beta m}K_{\beta \alpha
	m^{\prime }}+X_{m^{\prime }m}^{\ast }K_{\beta \alpha m}K_{\alpha \beta
	m^{\prime }}\right)\right]  \label{Inte}
\end{equation}
with \ $K_{\alpha\beta m}$ the generator $K_{\alpha\beta}$ of the $m$-th copy, $\chi_\alpha$ a set of $N$ free parameters and the matrices $X$ and
$Z$ defined as
\begin{equation}
X_{mm^{\prime }}=\frac{e^{\mathrm{i}\left( z_{m^{\prime }}-z_{m}\right) }}{\sin
	\left( z_{m^{\prime }}-z_{m}\right) }\text{ \ \ \ and \ \ \ }Z_{m^{\prime
	}m}=\cot \left( z_{m^{\prime }}-z_{m}\right),  \label{XZ}
\end{equation}
with $z_m$ a set of $M$ free parameters. The integrals of motion\ $R_m$ commute between themselves, $[R_{m},R_{m^{\prime}}]=0$, and with both the
quadratic Casimir  $C^2_m=\sum_{\alpha%
	\beta}^N K_{\alpha\beta m}K_{\beta\alpha m}$ and the linear Casimir $C^1_{m}=\sum_{\alpha}^N
K_{\alpha\alpha m}$ operators of each copy. They also commute with the conserved quantities
\begin{equation}
S_\alpha = \sum_{m=1}^M K_{\alpha\alpha m },
\label{eq:s_alpha_m_copies}
\end{equation}
corresponding to the operators\ \eqref{Cs} in the Liouvillian case of\ $M=2$.

The HW state of each copy, $\left\vert \Lambda
_{m}\right\rangle $ (the vacuum of the ladder operators $K_{\alpha \beta
	m}\left\vert \Lambda _{m}\right\rangle =0$ for all $\alpha <\beta $), is an eigenvector of the Cartan operator
\begin{equation}
K_{\alpha \alpha m}\left\vert \Lambda _{m}\right\rangle =\lambda_{\alpha
	m}\left\vert \Lambda _{m}\right\rangle.
\label{eq:cartan_operators}
\end{equation}
The eigenvalues $\lambda_{\alpha m}$ allow to characterize unambiguously the
$SU(N)$ \textit{irrep}  of the $m$-th copy and determine the
eigenvalues of the integrals of motion (\ref{Inte})
\begin{equation}
r_{m}=\sum_{\alpha =1}^{N}\chi _{\alpha }\lambda_{\alpha m}+\sum_{m^{\prime
	}\left( \neq m\right) =1}^{M}\cot \left( z_{m^{\prime }}-z_{m}\right)
\sum_{\alpha =1}^{N}\lambda_{\alpha m}\lambda_{\alpha
	m^{\prime }}+\sum_{\alpha =1}^{N}\left( \lambda_{\alpha m}-\lambda_{\alpha
	+1,m}\right) \sum_{i=1}^{M_{a}}\cot \left( z_{m}-E_{i}^{a}\right).
\label{Eigen}
\end{equation}
These eigenvalues depend also on $N-1$ sets of spectral parameters $%
E_{i}^{a} $ ($a=1,\cdots ,N-1$), which define every common eigenstate of the
integrals of motion (see Appendix \ref{app:wavefun}), and are obtained from particular
solutions of the sets of $N-1$ nonlinear coupled RG equations
\begin{equation}
\sum_{b=1}^{N-1}\sum_{i^{\prime }=1}^{M_{b}}{}^{\prime }A_{ba}\cot \left(
E_{i^{\prime }}^{b}-E_{i}^{a}\right) -\sum_{m=1}^{M}\left( \lambda_{a
	,m}-\lambda_{a +1,m}\right) \cot \left( z_{m}-E_{i}^{a}\right) =\chi
_{a}-\chi _{a+1},\qquad (a=1,\cdots,N-1)  \label{RGequ}
\end{equation}%
where the prime in the second sum indicates that the $i^{\prime }=i$ term is excluded
when $a=b$, and $A_{ab}=2\delta _{a,b}-\left( \delta
_{b,a-1}+\delta_{b,a+1}\right) $ is the $su\left( N\right) $ Cartan matrix
of dimension $\left( N-1\right) \times $ $\left( N-1\right)$.

The number of parameters $E_{i}^{a}$ within each set $a$ is given by
\begin{equation}
M_{a}=\sum_{\beta=1}^{a}\left[ \sum_{m=1}^{L}\lambda_{\beta m}-s_{\beta}\right],
\label{eq:m_alpha}
\end{equation}%
with $s_\beta$ the quantum numbers associated to the conserved quantities $S_\beta$ in \eqref{eq:s_alpha_m_copies}.

The set of integrals of motion (\ref{Inte}) with the matrices $X$ and $Z$ (%
\ref{XZ}), their eigenvalues (\ref{Eigen}) and eigenfunctions (Appendix \ref{app:wavefun}),
together with the set of RG equations (\ref{RGequ}) constitute the exact solution
of the $SU(N)$ trigonometric RG model. Any operator expressed
as a function of the integrals of motion is exactly solvable with
eigenvalues given by the same function of the eigenvalues of the integrals
of motion.

\section{Richardson-Gaudin models of dissipative $N$-level atom systems}

In this Section we show how the Liouvillian \eqref{L_RG} describing a
dissipative system of $L\,\ N$-level atoms can be obtained
from the trigonometric $SU(N)$ RG model. Since in the vectorized form the Liouvillian couples two $su(N)$ algebras, we restrict the RG model to a combination of two $su(N)$
copies ($M=2$), with $K_{\alpha \beta ,1}=K_{\alpha \beta }$ and $K_{\alpha
	\beta ,2}=J_{\alpha \beta }$. From the two $z_m$ parameters in the $X$ and $%
Z $ matrices we freely choose $z_{1}=0$, and $z_{2}=z\in \mathbb{C}$. With
these assumptions, the two integrals of motion reduce to
\begin{equation}
\begin{split}
R_{1} =& \sum_{\alpha =1}^{N}\chi _{\alpha }K_{\alpha \alpha ,1}+\cot
z\sum_{\alpha =1}^{N}K_{\alpha \alpha ,1}K_{\alpha \alpha ,2}+\sum_{\beta
	>\alpha =1}^{N}\frac{1}{\sin z}\left[ e^{\mathrm{i}z}K_{\alpha \beta ,1}K_{\beta
	\alpha ,2}+e^{-\mathrm{i}z}K_{\beta \alpha ,1}K_{\alpha \beta ,2}\right]  \\
R_{2} =& \sum_{\alpha =1}^{N}\chi _{\alpha }K_{\alpha \alpha ,2}-\cot
z\sum_{\alpha =1}^{N}K_{\alpha \alpha ,1}K_{\alpha \alpha ,2}-\sum_{\beta
	>\alpha =1}^{N}\frac{1}{\sin z}\left[ e^{\mathrm{i}z}K_{\alpha \beta ,1}K_{\beta
	\alpha ,2}+e^{-\mathrm{i}z}K_{\beta \alpha ,1}K_{\alpha \beta ,2}\right].
\end{split}
\label{eq:integrals_of_motion}
\end{equation}
In order to establish the correspondence with $\mathcal{L}_{RG}$ \eqref{L_RG}, we consider
the linear combination
\begin{equation}
\begin{split}
g\sin z~\left( R_{1}-R_{2}\right) =&\ g\sin z\sum_{\alpha =1}^{N}\chi _{\alpha }\left( K_{\alpha \alpha,1}-K_{\alpha \alpha ,2}\right) \\
&+2g\cos z\sum_{\alpha =1}^{N}K_{\alpha	\alpha ,1}K_{\alpha \alpha ,2} +2g\sum_{\beta >\alpha }^{N}\left[e^{\mathrm{i}z}K_{\alpha \beta ,1}K_{\beta \alpha ,2} + e^{-\mathrm{i}z}K_{\beta \alpha,1}K_{\alpha \beta ,2}\right].
\end{split}
\label{Linear}
\end{equation}
Comparing the last two terms in the equation above with those of $\mathcal{L}%
_{RG}$ (\ref{L_RG}), we get the following relations between the
RG-model parameters and those of the Liouvillian
\begin{equation}
2ge^{\mathrm{i}z}=-\Gamma \left( 1+p\right) \text{ \ \ \ \ \ and \ \ \ \ }%
2ge^{-\mathrm{i}z}=-\Gamma \left(1-p\right).  \label{eq2}
\end{equation}
By adding these two relations we get also the correspondence between the
second terms in (\ref{Linear}) and $\mathcal{L}_{RG}$, $2g\cos z=-\Gamma $, while the difference gives $2g\sin z = \mathrm{i}\Gamma p$. The ratio of the latter equalities defines the parameter $z$ in terms of the
polarization parameter in the Liouvillian
\begin{equation}
\cot z=\frac{\mathrm{i}}{p}.  \label{z}
\end{equation}
By taking the product of the equalities \eqref{eq2}, we obtain the
relation between the parameter $g$ in the linear combination (\ref%
{Linear}) and the Liouvillian parameters
\begin{equation}
g=\frac{\Gamma \sqrt{1-p^{2}}}{2} .  \label{g}
\end{equation}
We still have to determine the parameters $\chi_\alpha$ in the linear term
of equation (\ref{Linear}). Comparison with the linear term in $\mathcal{L}_{RG}$ results in  $\chi_{\alpha }\,g\sin z = \frac{\Gamma p}{2}\left( N+1-2\alpha \right)$, and using
the previous relations
\begin{equation}
\chi _{\alpha }=\mathrm{i}\left( 2\alpha -N-1\right).  \label{chi}
\end{equation}
We have determined the parameters of the RG\ integrals of motion $z$ (\ref{z}%
), $\chi _{\alpha }$ (\ref{chi}) and the parameter $g$ (\ref{g}) appearing
in the linear combination (\ref{Linear}), thus establishing the
correspondence between the RG integrals of motion and the Liouvillian (%
\ref{L_RG}).


In order to obtain the eigenvalues of $\mathcal{L}_{RG}$ written as a linear combination of the integrals of motion \eqref{eq:integrals_of_motion} we
have to establish the HW states of the two $su\left( N\right) $ copies. For
the first copy, the HW state is $\left\vert \Lambda _{1}\right\rangle
=\left\vert L,0,\cdots ,0\right\rangle $. The eigenvalues of the Cartan
operators \eqref{eq:cartan_operators} for this state are $\lambda_{\alpha 1} = \delta_{\alpha,1}L$. For the second copy, we
have performed the transformation $J_{\alpha \beta }=-\overline{J}_{\beta
	\alpha }$, which inverts the HW state to $\left\vert \Lambda
_{2}\right\rangle =\left\vert 0,\cdots ,0,L\right\rangle $ with eigenvalues $\lambda_{\alpha 2}=-\delta_{\alpha,N} L$. Having determined the labels $\lambda_{\alpha m}$, the eigenvalues of the two integrals of motion \eqref{Eigen} are
\begin{equation}
\begin{split}
r_{1} =&-\mathrm{i}\left( N-1\right) L-L\sum_{i=1}^{M_{1}}\cot E_{i}^{\left(
	1\right) } \\
r_{2} =&-\mathrm{i}\left( N-1\right) L+L\sum_{i=1}^{M_{N-1}}\cot \left(
z-E_{i}^{\left( N-1\right) }\right) .
\end{split}
\label{EigenInt}
\end{equation}
Therefore, the eigenvalues of the RG part of the Liouvillian are obtained
from the linear combination (\ref{Linear}) as
\begin{equation}
l_{RG}=g \sin z\left( r_{1}-r_{2}\right)=-\mathrm{i}\frac{L\Gamma p}{2}\left[
\sum_{i=1}^{M_{1}}\cot E_{i}^{\left( 1\right) }+\sum_{i=1}^{M_{N-1}}\cot
\left( z-E_{i}^{\left( N-1\right) }\right) \right] .  \label{lrg}
\end{equation}
Finally, the eigenvalues of the complete Liouvillian are
\begin{equation}
l=l_C+l_{RG}=-\mathrm{i}\sum_{\alpha}\varepsilon_\alpha s_\alpha-\Gamma
\left(L^2+(N-1)L\right)+\frac{\Gamma-\Gamma_0}{2}\sum_{\alpha}s_%
\alpha^2+l_{RG}.  \label{lTot}
\end{equation}
The spectral parameters $E_i^a$ are determined by the RG equations (%
\ref{RGequ}), which, in this particular case, reduce to
\begin{equation}
\begin{split}
&\sum_{b =1}^{N-1}\sum_{i^{\prime }=1}^{M_{b }}{}^{\prime }A_{b 1 }\cot
\left( E_{i^{\prime }}^{\left( b \right) }-E_{i}^{\left( 1 \right) }\right) + L\cot \left( E_{i}^{\left( 1\right) }\right)=-2\mathrm{i} \\
& \sum_{b =1}^{N-1}\sum_{i^{\prime }=1}^{M_{b }}{}^{\prime }A_{b a }\cot
\left( E_{i^{\prime }}^{\left( b \right) }-E_{i}^{\left( a \right) }\right)=-2\mathrm{i},\qquad\qquad\qquad\qquad (a =2,\cdots ,N-2) \\
&\sum_{b =1}^{N-1}\sum_{i^{\prime }=1}^{M_{b }}{}^{\prime }A_{b N-1 }\cot
\left( E_{i^{\prime }}^{\left( b \right) }-E_{i}^{\left( N-1 \right) }\right) - L\cot \left( z-E_{i}^{\left( N-1\right) }\right)=-2\mathrm{i},
\end{split}
\label{eq:RESUN}
\end{equation}
with the prime in the sum terms as defined in \eqref{RGequ}.

The number of spectral parameters for each set $E_{i}^{(a)}$ ($a =1,\cdots,
N-1)$ \eqref{eq:m_alpha} is
\begin{equation}
M_{a }=L-\sum_{\beta=1}^{a}s_\beta.
\end{equation}
The SS belongs to the sector with $s_\alpha=0$, for which the
number of spectral parameters is $M_{a}=L$ for all $a$.

\section{$SU(2)$ Richardson-Gaudin: a dissipative collective spin}

A system of two-level atoms coupled to an environment is the simplest
application of the previous general solution. The algebra associated to this
simplest case is the rank-one $su(2)$. The exact solution has been
previously obtained in \cite{Rib-Pro}, although the authors followed a
different approach based on coherent states and treated the $SU(2)$ operators
as a collective spin. We will demonstrate that the solution in \cite{Rib-Pro} can also be interpreted as an open system of two-level atoms
and that, except for a term depending on the  quantum numbers $s_\alpha$, it can be
obtained as the limit $SU(2)$ of the general $SU(N)$ RG
integrable Liouvillians derived above.

In their work, Ribeiro and Prosen studied the dissipative dynamics
of a collective spin $s$ under a local field $h$
\begin{equation}
H=-hS_{z}
\end{equation}%
coupled to an environment characterized by the jump operators
\begin{equation}
W_{0}=\sqrt{4\Gamma _{0}}S_{z},\quad W_{\pm }=\sqrt{\Gamma \left( 1\mp
	p\right) }S_{\pm }\text{, }
\end{equation}%
note that we have rescaled parameters $\Gamma$ and $\Gamma_0$ of \cite{Rib-Pro} for ease of consistency with our general $SU(N)$ models. In the
vector representation the Ribeiro-Prosen Liouvillian in terms of two
collective spins takes the form
\begin{equation}
\begin{split}
\mathcal{L}_{RP}=&\ \mathrm{i}h (S_z-\bar{J}_z)+4\Gamma_0\left[S_z \bar{J}_z-\frac{1}{%
	2}\left(S_z^2+\bar{J}_z^2\right)\right]  \\
&+\Gamma(1-p) \left[S_+ \bar{J}_+-\frac{1}{2}\left(S_-S_+ + \bar{J}_-\bar{J%
}_+\right)\right] +\Gamma(1+p) \left[S_- \bar{J}_--\frac{1}{2}\left(S_+S_- +
\bar{J}_+\bar{J}_-\right)\right].
\end{split}
\label{LRP}
\end{equation}
By expressing the spin operators as
\begin{eqnarray}
S_z=\frac{K_{22}-K_{11}}{2} & \ \ \ & \overline{J}_z=\frac{\overline{J}_{22}-%
	\overline{J}_{11}}{2}=\frac{J_{11}-J_{22}}{2}  \notag \\
S_+=K_{21}& \ \ \ &\bar{J}_+=\bar{J}_{21}=-J_{12} \\
S_-=K_{12}& \ \ \ & \bar{J}_-=\bar{J}_{12}=-J_{21},  \notag
\end{eqnarray}
and fixing the parameters $\varepsilon_1=-h/2$ and $\varepsilon_2=h/2$, it
is straightforward to establish the relation with the general $SU(N)$
Liouvillian of equations \eqref{L_C} and \eqref{L_RG} in the $SU(2)$ limit
\begin{equation}
\mathcal{L}_{RP}=\mathcal{L}_{SU(2)}+\Gamma_0\, s_1s_2,
\end{equation}
We note that the spectrum of both Liouvillians coincide in the subspace $s_{1}=s_{2}=0$ which contains the SS. The other sectors are displaced by a constant
depending on the quantum numbers $s_{1}$ and $s_{2}$.

The Liouvillian eigenvalues \eqref{lTot}, $l_{SU(2)}=l_C+l_{RG}$, of $\mathcal{L}_{SU(2)}$ are
\begin{equation}
l_{SU(2)}=-\mathrm{i}h s_1-\Gamma(L^2+L)+s_1^2(\Gamma-\Gamma_0)
-\frac{\mathrm{i}L \Gamma p}{2}\sum_{i=1}^{M_1}\left[\cot(E_i)+\cot(z-E_i)
\right],
\label{LSU2}
\end{equation}
with $\cot(z)=\mathrm{i}/p$. Since $su(2)$ is an algebra of rank one, only one set of spectral parameters $%
E_i$  appear in the solution. The number of
spectral parameters is given by $M_1=L-s_1$, and they are determined by a
unique set of RG equation \eqref{eq:RESUN}
\begin{equation}
2\sum_{i^{\prime}\not=i}^{M_1}\cot(E_{i^{\prime}}-E_i)+L\cot(E_i)-L%
\cot(z-E_i)=-2 \mathrm{i}.
\label{RGSU2}
\end{equation}
We have, therefore, demonstrated the integrability of $\mathcal{L}_{RP}$ (\ref{LRP}) by expressing it as a linear combination of the complete set of integrals of motion of the $SU(2)$ trigonometric RG model. Moreover, we derived the Liouvillian eigenvalues (\ref{LSU2}) and the RG equations (\ref{RGSU2}) that determine the spectral parameters $E_i$, completing in this way the exact solution of the dissipative collective spin model.

\section{$SU(3)$ Richardson-Gaudin models: dissipative 3-level atoms}

The next example in order of complexity is the case of three-level atoms
that we will study in this Section. This case is associated with  the rank-two $su(3)$ algebra. After discussing the RG equations and its
equivalence to an electrostatic-like problem, we present exact solutions for the SS and several slow decaying modes.

\subsection{Exact solution}
For $N=3$, we have two sets of spectral
parameters $E_{i}^{\left(1\right)}$ and $E_{i}^{\left(2\right)}$ that fulfill the two sets of RG equations
\begin{equation}
\begin{split}
2\sum_{i^{\prime }\neq i=1}^{M_{1}}\cot \left( E_{i^{\prime }}^{\left(
	1\right) }-E_{i}^{\left( 1\right) }\right) -\sum_{i^{\prime }=1}^{M_{2}}\cot
\left( E_{i^{\prime }}^{\left( 2\right) }-E_{i}^{\left( 1\right) }\right)
+L\cot \left( E_{i}^{\left( 1\right) }\right) =&\ -2\mathrm{i}\\
2\sum_{i^{\prime }\neq i=1}^{M_{2}}\cot \left( E_{i^{\prime }}^{\left(
	2\right) }-E_{i}^{\left( 2\right) }\right) -\sum_{i^{\prime }=1}^{M_{1}}\cot
\left( E_{i^{\prime }}^{\left( 1\right) }-E_{i}^{\left( 2\right) }\right)
-L\cot \left( z-E_{i}^{\left( 2\right) }\right) =&\ -2\mathrm{i},
\end{split}
\end{equation}
where the number of parameters $E_i^{(1)}$ and $E_{i}^{(2)}$ is given by $M_1=L-s_1$ and $M_2=L+s_3$, respectively. These equations can be rewritten as
\begin{equation}
\begin{split}
\sum_{i^{\prime }\neq i=1}^{M_{1}}\frac{2}{e_{i}-e_{i^{\prime }}}%
-\sum_{i^{\prime }=1}^{M_{2}}\frac{1}{e_{i}-\omega _{i^{\prime }}}+\frac{%
	Q_{+}^{e}}{e_{i}-\mathrm{i}} + \frac{Q_{-}^{e}}{e_{i}+\mathrm{i}} &= 0 \\
\sum_{i^{\prime }\neq i=1}^{M_{2}}\frac{2}{\omega _{i}-\omega _{i^{\prime }}}%
-\sum_{i^{\prime }=1}^{M_{1}}\frac{1}{\omega _{i}-e_{i^{\prime }}}+\frac{%
	Q_{+}^{\omega }}{\omega _{i}-\mathrm{i}}+\frac{Q_{-}^{\omega }}{\omega _{i}+\mathrm{i}}-\frac{L%
}{\omega _{i}-\frac{\mathrm{i}}{p}} &= 0
\end{split}
\label{eq:rg_equations_su3}
\end{equation}
with $e_{i}=\cot E_{i}^{\left( 1\right) }$, $\omega _{i}=\cot
E_{i}^{\left(2\right) }$, and the effective charges $Q_\pm^{e,\omega}$
\begin{equation}
\begin{split}
Q_{+}^{e}&=2+\frac{s_1-s_2}{2},\qquad Q_{-}^{e}=\frac{s_1-s_2}{2}\\
Q_{+}^{\omega }&=2+\frac{s_2-s_3}{2},\qquad Q_{-}^{\omega }=%
\frac{s_2-s_3}{2}.
\end{split}
\end{equation}
Equations \eqref{eq:rg_equations_su3} can be interpreted as a classical
electrostatic-like problem in two dimensions for two classes of particles,\ $e_{i}$ and\ $\omega_{i}$,
with opposite charges. Particles of the same class repel each
other with an effective strength twice as large as the one with which distinct kind of particles are attracted. Moreover, depending on the sign of the effective charges, the\ $e_i$ feel the repulsion or attraction  of charges $Q_{+}^{e}$ fixed at position $\mathrm{i}$ and $Q_{-}^{e}$ at position $-\mathrm{i}$. Similarly, the $\omega_i$ are repelled or attracted  by charges $Q_{+}^{\omega }$ at position $\mathrm{i}$ and $Q_{-}^{\omega }$ at position $-\mathrm{i}$ and, additionally, they are  attracted by a charge of magnitude $L$ located at
position $\frac{\mathrm{i}}{p}$. The solutions of these equations provide
the equilibrium positions of both type of particles, which  determine
 the eigenvectors (see Appendix \ref{app:wavefun}) as well as the Liouvillian eigenvalues \eqref{lTot}
\begin{equation}
l=-\mathrm{i}\sum_{\alpha =1}^{3}\varepsilon _{\alpha }s_\alpha-\Gamma \left( L^{2}+2L\right) +\frac{\Gamma -\Gamma
_{0}}{2}\sum_{\alpha =1}^{3}s_\alpha ^{2} -\mathrm{i}%
\frac{L\Gamma p}{2}\left[ \sum_{i=1}^{M_{1}}e_{i}+\sum_{i=1}^{M_{2}}\frac{%
\mathrm{i}\omega _{i}+p}{p\omega _{i}-\mathrm{i}}\right].  \label{lSU3}
\end{equation}

\subsection{Steady state as a function of $p$}

As a first example, we solve the $SU(3)$ RG equations for the SS of a system with $L=40$ atoms. The SS belongs to the sector $s=(s_1,s_2,s_3)=(0,0,0)$, implying that the number of spectral parameters is
$M_{1}=M_{2}=L$ and the charges become $Q_{+}^{e}=Q_{+}^{\omega }=2$ and $%
Q_{-}^{e}=Q_{-}^{\omega }=0$. Note that the only parameter of the
Liouvillian entering in the RG equations \eqref{eq:rg_equations_su3} is the polarization factor $p$ that defines the position $\mathrm{i}/p$ of the effective charge $L$, which strongly attracts the spectral parameters
 $\omega_i$. The position of this
effective charge and that of charges $Q_{+}^{e},\ Q_{+}^{\omega }$
located at $+\mathrm{i}$,  together with the mutual repulsion or attraction between the spectral parameters determine their electrostatic equilibrium position.

\begin{figure}[t]
	\includegraphics[width=.5\textwidth]{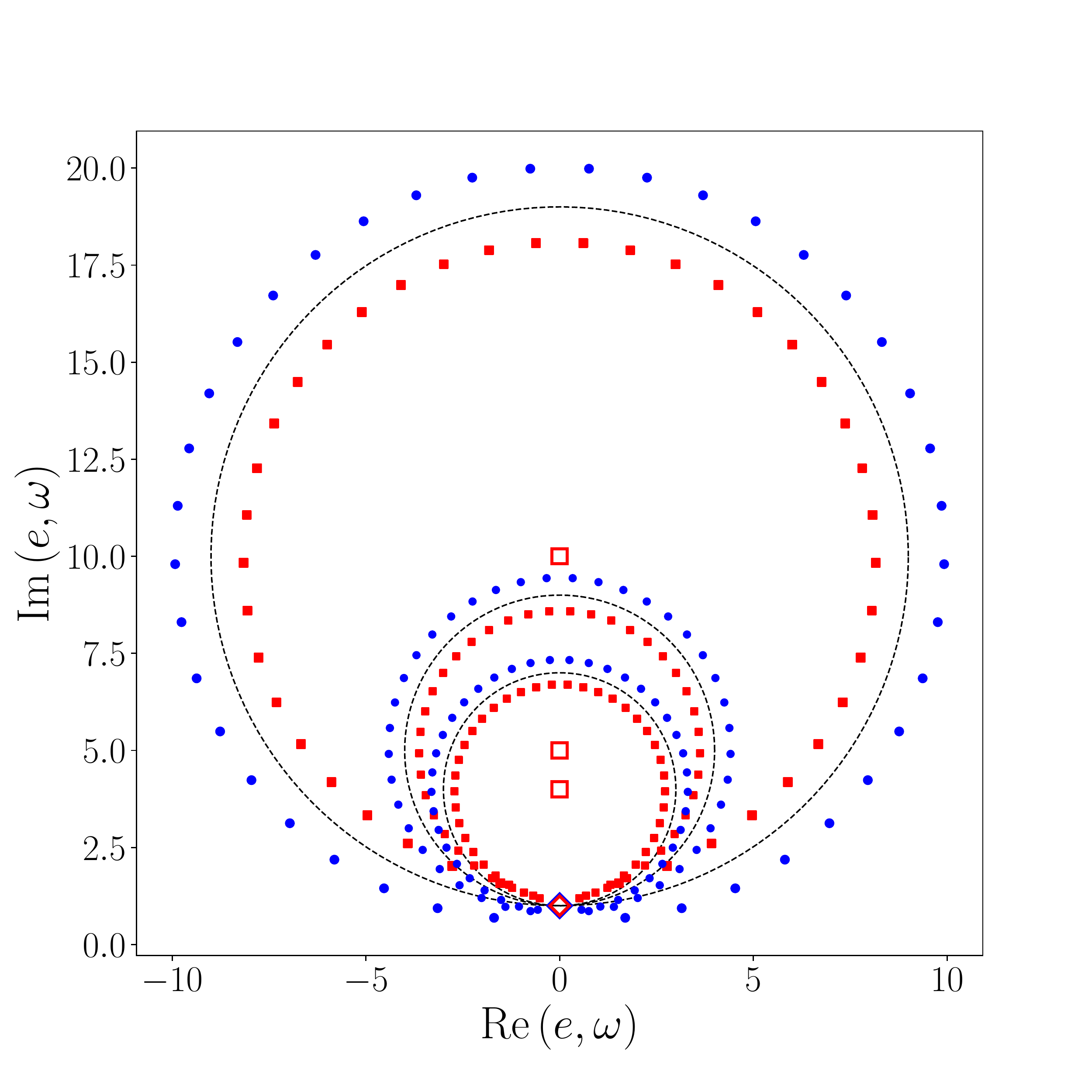}
	\caption{Spectral parameters\ $e_i$ (blue circles) and\ $\protect\omega_i$ (red
		squares) of the exact solutions of the RG equations \eqref{eq:rg_equations_su3} for the SS in a system of $L=40$ atoms and different
		values of $p$. The dashed lines are circles with radii\ $|1-1/p|$ centered
		at the position $\mathrm{i}/p$ of the charge $L$ (red squares). Blue and red diamonds show the position, $+\mathrm{i}$, of repulsive charges $Q_{+}^{e}$ and $Q_{+}^{\omega}$ (see text). Data  shown for $p=0.1,\ 0.2,\ 0.25$. }
	\label{fig:exact_steady_state}
\end{figure}

Figure \ref{fig:exact_steady_state} shows the equilibrium position of the $\omega_i$ and $e_i$ parameters of the SS for different values of\ $p$. As it can be seen, the spectral parameters distribute close to a circle with
center at the position $\mathrm{i}/p$ of charge $L$ whose radius is determined by the
distance between $\mathrm{i}/p$ and the position of charges $%
Q_{+}^{e,\omega}$, $r=|\mathrm{i}/p-\mathrm{i}|=|1-1/p|$. The $\omega_i$ parameters
accommodate inside the circle equilibrating the attraction of charge $L$ and
that of parameters $e_i$, which are located outside the circle. Both types of parameters spread almost uniformly within their respective arcs due to their electrostatic repulsion. However, they are weakly  disturbed by the charges $Q_{+}^{e,\omega}$, of small magnitude, $2$, as compared to $L$,  producing a depletion close to their position $+\mathrm{i}$. We have
verified, by substituting the values of the spectral parameters in \eqref{lSU3}, that the eigenvalue of the SS is exactly zero.

We note that the distribution of the spectral parameters in the complex plane
completely determines the eigenvalue and eigenvector of the  Liouvillian SS. This distribution is universal in the sense that it does not
depend on the parameters $\Gamma$ and $\Gamma_0$ and it just scales with $p$, which determines the center and radius of the circle. For $p\rightarrow 1$
the circle shrinks and the whole set of parameters $e_i$ and $\omega_i$
collapses to $+\mathrm{i}$, whereas as $p$ approaches to zero the circle's
radius and its center increase boundlessly, so that the modulus of several spectral parameters becomes infinity.

\subsection{Full spectrum of the Liouvillian}

\begin{figure}[t]
\includegraphics[width=.86\textwidth]{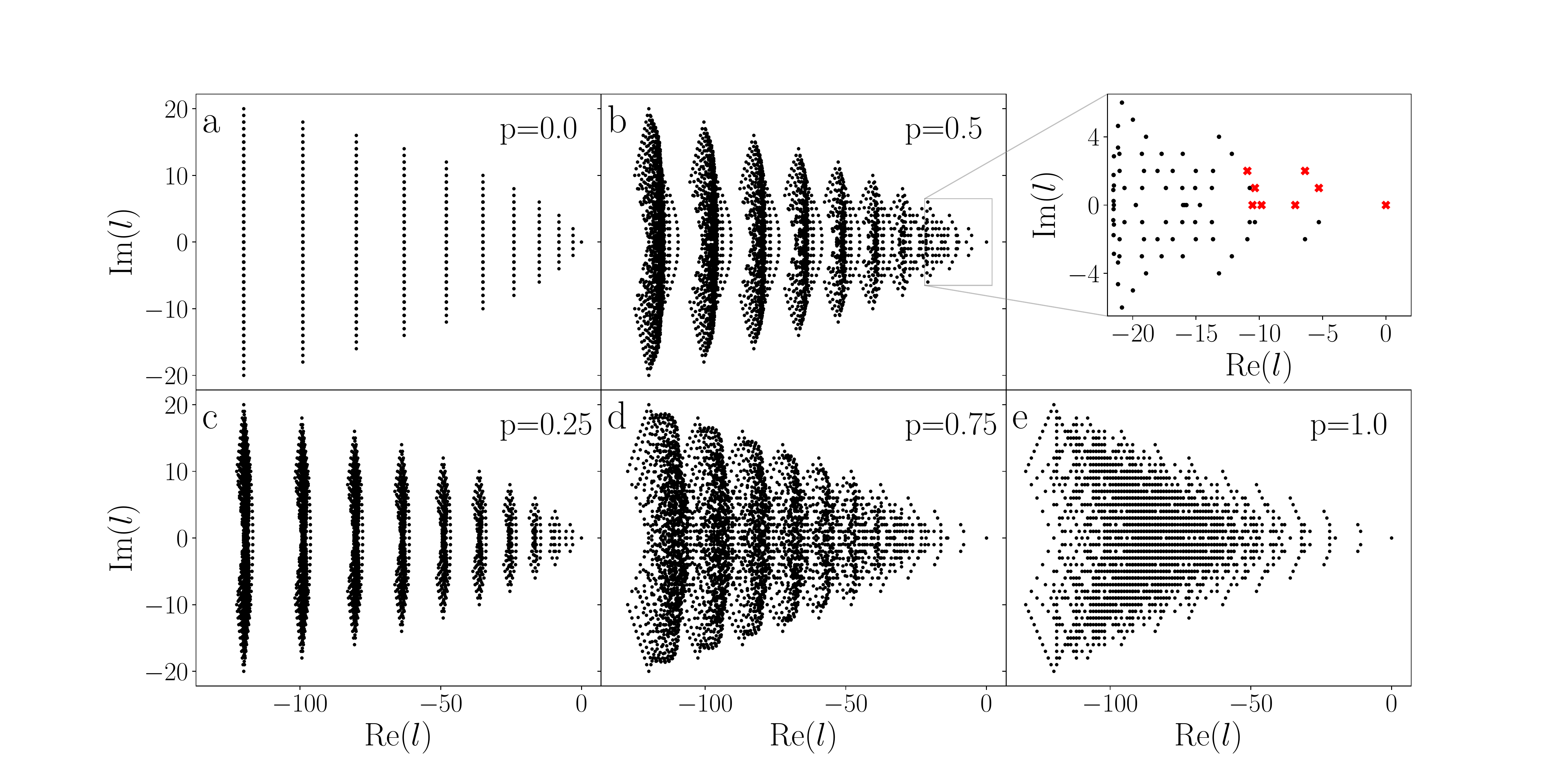}
\caption{Spectrum of  the $SU(3)$ Liouvillian for different values of\ $p$. Data
shown for $L=10,\ \Gamma =\Gamma _{0}=1,$ and  $\protect\varepsilon=(\varepsilon_1,\varepsilon_2,\varepsilon_3)=(-1,0,1)$.
For $p=0.5$ we show a zoom of the spectrum near the region\ $\text{Re}(%
\protect l) = 0$. Red crosses mark selected eigenstates whose spectral parameters are shown
in figure \ref{fig:exact_solution}.}
\label{fig:liouvillian_spectra}
\end{figure}

The general $SU(3)$ Liouvillian commutes with the\ $S_{\alpha}$ operators \eqref{Cs}, grouping its eigenvalues into sectors labeled by
the quantum numbers\ $(s_1,s_2,s_3)$. All eigenvalues belonging to the same sector organize into horizontal lines in the complex plane (see  figure~\ref{fig:liouvillian_spectra}) with
 constant imaginary part given by  $-\mathrm{i}\sum_{\alpha=1}^{3}\varepsilon_{\alpha } s_{\alpha}$. The  possible allowed values of integer numbers $(s_1,s_2,s_3)$ are given by $-L\leq s_i\leq L$ with the constraint $s_1+s_2+s_3=0$. Since for every set $(s_1,s_2,s_3)$ there exists
the conjugate $(-s_1,-s_2,-s_3)$, the eigenvalues appear in complex conjugate pairs.
The SS belongs to the sector\ $s=(0,0,0)$ and the
dissipative gap is given by an eigenvalue located in one of the sectors with\ $\max \left\lbrace s_\alpha \right\rbrace = 1$.

We show in figure~\ref{fig:liouvillian_spectra} the full spectrum of several $SU(3)$
Liouvillians for different values of\ $p$ and $L=10$. Due to the small value of $L$ we could use an exact diagonalization procedure explained in Appendix \ref{app:exact_diagonalization}. We observe that, for $p\approx 0$, the eigenvalues cluster in groups with similar real values, and that the number of eigenvalues in these groups increases as the real value becomes more negative.  The origin of this additional  grouping can be traced back to the conservation of the total $SU(3)$  symmetry for $p=0$. In this limit  the Liouvillian can be written in terms of the total $SU(3)$ quadratic Casimir operator as

\begin{equation}
\begin{split}
\mathcal{L} = -\mathrm{i}\sum_{\alpha }\varepsilon _{\alpha }S_\alpha  +%
\frac{\Gamma -\Gamma _{0}}{2}\sum_{\alpha }S_\alpha^{2} -\Gamma \frac{C^2_T}{2},
\end{split}%
\label{p0}
\end{equation}
where $C^2_T=\sum_{\alpha\beta} \left(K_{\alpha\beta}+J_{\alpha\beta}\right)\left(K_{\beta\alpha}+J_{\beta\alpha}\right)$. To obtain the eigenvalues of the total Casimir operator, we  consider the $SU(3)$ couplings of the two $SU(3)$ irreducible representations: $(L,0)\otimes(0,L)=\bigoplus_{\lambda=0}^L  (\lambda,\lambda)$. For the  multiplets  $(\lambda,\lambda)$,   the eigenvalues of the total Casimir are  $c_T^2=2(\lambda^2+ 2\lambda)$.  For $\Gamma=\Gamma_0$, these values determine the real part of the spectrum, as it is shown in figure \ref{fig:liouvillian_spectra}(a). The imaginary part is given by the first term in (\ref{p0}) that depends on the configurations ($s_1,s_2,s_3$) and the level energies $\varepsilon_{\alpha}$. Every set of eigenvalues with the same real part corresponds to the total $SU(3)$ multiplet  $(\lambda,\lambda)$, whose dimension, $(\lambda+1)^3$, increases with $\lambda=0,\cdots,L$. In particular, the SS corresponds to the singlet $(0,0)$. For $p \neq 0$ the total $SU(3)$ symmetry is broken and its different multiplets mix. However, as it can be seen in figures \ref{fig:liouvillian_spectra}(b-d), the eigenvalues still preserve the band structure of $p=0$ up to the extreme limit of $p=1$, shown in figure\ \ref{fig:liouvillian_spectra}(e).

\begin{figure}[t]
\includegraphics[width=.76\textwidth]{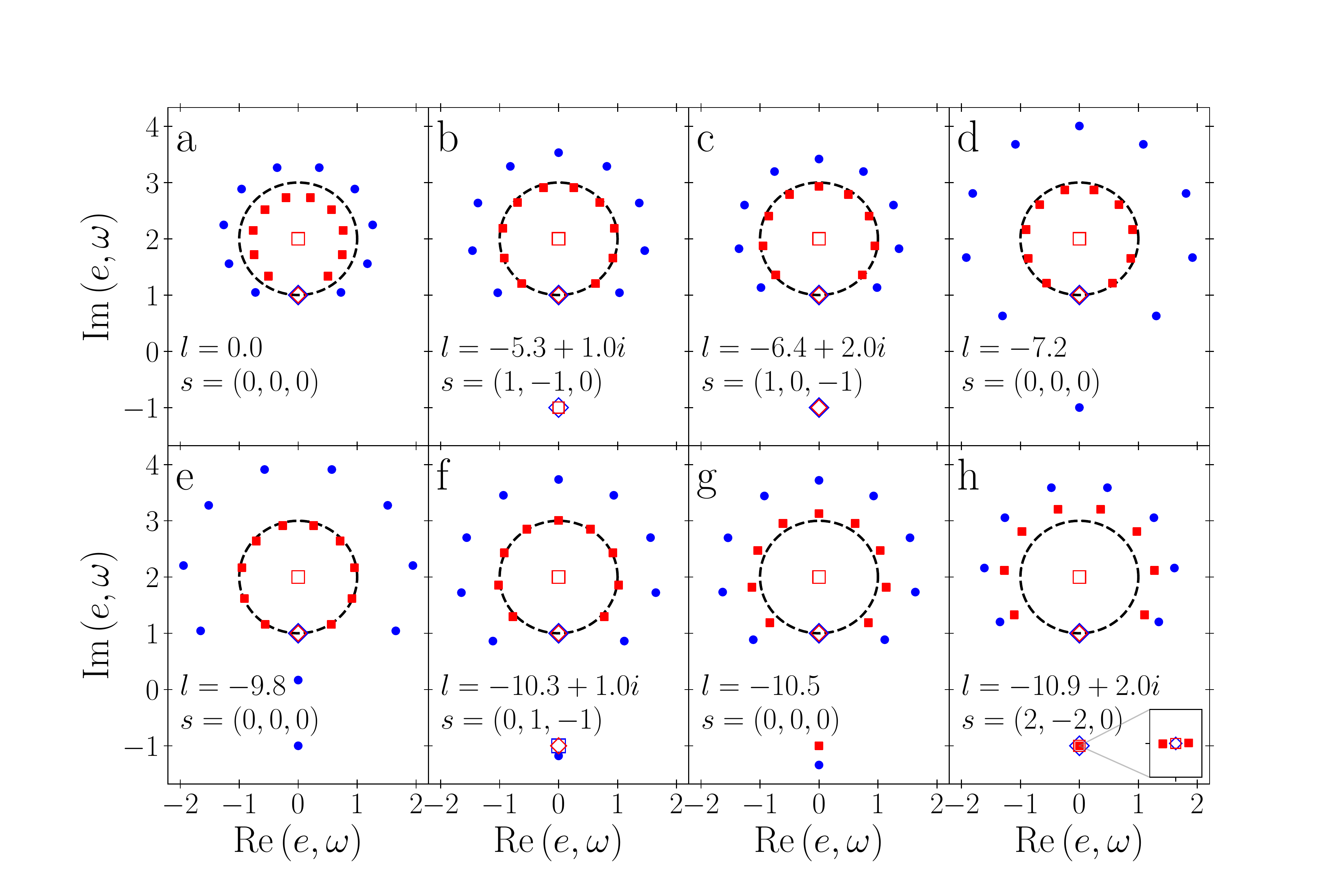}
\caption{Spectral parameters\ $e_i$ (blue circles) and\ $\protect\omega_i$ (red
squares) of the exact solution of the RG equations \eqref{eq:rg_equations_su3} for the SS and slowest decaying eigenstates highlighted in figure \ref{fig:liouvillian_spectra}(b, inset). The dashed lines represent a circle with radius\
$|1-1/p|$ centered at the position $\mathrm{i}/p$ (red square). Eigenvalues\ $l$ and symmetry quantum numbers $s=(s_1,s_2,s_3)$ are indicated in each panel. Open symbols represent the position of the effective charges, $Q_{\pm}^{e,\omega}$ and $L$ of the electrostatic-like equivalent problem (see text). The color (blue or red) of the open symbols indicates the kind of spectral parameters with which the charges interact (parameters $e_i$ or $\omega_i$), while  their form indicates if the charge is attractive (squares) or repulsive (diamonds). Data shown for: $L=10,\
p=0.5,\ \Gamma = \Gamma_{0} = 1,\ \protect\varepsilon = (-1,0,1)$.}
\label{fig:exact_solution}
\end{figure}

The eigenvalues  of the Liouvillian can be examined on the light of the exact solution. In figure~\ref{fig:exact_solution} we show, for $p=0.5$, the solution of  the RG
equations \eqref{eq:rg_equations_su3} for the SS and seven of the closest decaying states indicated by the red crosses in figure~\ref{fig:liouvillian_spectra}(b, inset). Each panel shows the eigenvalue and the quantum numbers\ $s_\alpha$ that fix the number of spectral parameters \ $%
M_{1}=L-s_1,\ M_{2}= L+s_3$ and the value of the charges $Q_{\pm}^{e,\omega}$.
For the SS and the two slowest decaying states of figures~\ref{fig:exact_solution} (a-c), all spectral parameters are distributed close to the circle of radius\ $r=|1-1/p|$ with center at\ $\mathrm{i}/p$, whereas for the other decaying modes, some spectral parameters leave the region of the circle and sit close to the position $-\mathrm{i}$, revealing a subtle detailed balance between the different charges. For example, in panel h two complex
 $\omega_i$ parameters surround the charge $Q_-^{\omega}$ which for this set of quantum numbers $s_{\alpha}$ is attractive ($Q_-^{\omega}=-1$).
 Since the repulsion strength between the like particles $\omega_i$ is twice the magnitude of the attraction exerted by the charge $Q_-^{\omega}=-1$ equidistantly located between them, we conclude that this subsystem close to the position $-\mathrm{i}$ is in electrostatic equilibrium. A similar analysis could be performed for each of the panels, making it possible to understand the position of the spectral parameters in the complex plane in terms of the global and local electrostatic equilibrium.

\section{Thermodynamic limit of the N-level atom system}

In this Section we calculate the wave function of the SS and the dissipative gap in the TL, $L\rightarrow \infty$. We start with the general Liouvillian \eqref{Liuv}, which upon the restriction (\ref{restriction}) reads
\begin{multline}
\mathcal{L}= -i\sum_{\alpha }\varepsilon _{\alpha }\left( K_{\alpha \alpha
}-\overline{J}_{\alpha \alpha }\right) -\Gamma C^{2}+\frac{\Gamma -\Gamma
	_{0}}{2}\sum_{\alpha }\left( K_{\alpha \alpha }-\overline{J}_{\alpha \alpha
}\right) ^{2} \\
 +\Gamma \sum_{\alpha }K_{\alpha \alpha }\overline{J}_{\alpha \alpha
}+\Gamma \left( 1-p\right) \sum_{\alpha >\beta }K_{\alpha \beta }\overline{J}%
_{\alpha \beta }+\Gamma \left( 1+p\right) \sum_{\alpha <\beta }K_{\alpha
	\beta }\overline{J}_{\alpha \beta }+\frac{\Gamma p}{2}\sum_{\alpha }\left(
N+1-2\alpha \right) \left( K_{\alpha \alpha }+\overline{J}_{\alpha \alpha
}\right).
\label{LL}
\end{multline}%
In order to work out the TL we will make use of the
Schwinger boson mean-field theory \cite{Kaup, Arovas}, which we extend to non-hermitian operators. The Schwinger boson mapping of the $SU(N)$
generators is
\begin{equation}
K_{\alpha \beta }=a_{\alpha }^{\dagger }a_{\beta },\qquad\overline{J}%
_{\alpha \beta }=b_{\alpha }^{\dagger }b_{\beta },
\label{Sch1}
\end{equation}%
with the restrictions
\begin{equation}
\sum_{\alpha =1}^{N}a_{\alpha }^{\dagger }a_{\alpha }=\sum_{\alpha
	=1}^{N}b_{\alpha }^{\dagger }b_{\alpha }=L.
\label{Sch2}
\end{equation}
Substituting the boson mapping into (\ref{LL}) and rescaling the dissipation strengths with
the size of the system, $\gamma =\Gamma \,L$ and $\gamma _{0}=\Gamma _{0}\,L$,
the Liouvillian is
\begin{align}
\mathcal{L}=& -\sum_{\alpha =1}^{N}\left( i\varepsilon _{\alpha }+\mu_a
\right) a_{\alpha }^{\dagger }a_{\alpha }+\sum_{\alpha =1}^{N}\left(
i\varepsilon _{\alpha }-\mu_b \right) b_{\alpha }^{\dagger }b_{\alpha }-\frac{%
	\gamma }{L}C^{2}+\frac{\gamma -\gamma _{0}}{2L}\sum_{\alpha }\left(
a_{\alpha }^{\dagger }a_{\alpha }-b_{\alpha }^{\dagger }b_{\alpha }\right)
^{2}
\label{eq:liouv_schwinger_rep}\\
& +\frac{\gamma }{L}\sum_{\alpha }a_{\alpha }^{\dagger }a_{\alpha }b_{\alpha
}^{\dagger }b_{\alpha }+\frac{\gamma }{L}\left( 1-p\right) \sum_{\alpha
	>\beta }a_{\alpha }^{\dagger }a_{\beta }b_{\alpha }^{\dagger }b_{\beta }+%
\frac{\gamma }{L}\left( 1+p\right) \sum_{\alpha <\beta }a_{\alpha }^{\dagger
}a_{\beta }b_{\alpha }^{\dagger }b_{\beta }+\frac{\gamma p}{2L}\sum_{\alpha }\left( N+1-2\alpha \right) \left(
a_{\alpha }^{\dagger }a_{\alpha }+b_{\alpha }^{\dagger }b_{\alpha }\right),
 \nonumber
\end{align}
where we have added chemical potentials $\mu _{a/b}$ to preserve the total
number of particles in each $SU(N)$ system.

In the TL we assume a boson
coherent state for the SS. Considering the non-hermiticity of the
Liouvillian, bras and kets are different
\begin{equation}
\begin{split}
|\Psi \rangle =& \exp \left[ \sqrt{L}\sum_{\alpha =1}^{N}\left( A_{\alpha
}a_{\alpha }^{\dagger }+B_{\alpha }b_{\alpha }^{\dagger }\right) \right]
|0\rangle  \\
\langle \overline{\Psi }|=& \langle 0|\exp \left[ \sqrt{L}\sum_{\alpha
	=1}^{N}\left( \overline{A}_{\alpha }a_{\alpha }+\overline{B}_{\alpha
}b_{\alpha }\right) \right],
\end{split}%
\end{equation}%
with\ $A_{\alpha },\overline{A}_{\alpha },B_{\alpha },\overline{B}_{\alpha }$
a set of parameters that we choose so that the expectation value\ $\langle
\overline{\Psi }|\mathcal{L}|\Psi \rangle $ is zero in the TL. The coherent bra and ket are the vacua of a new set
of shifted boson operators
\begin{equation}
c_{\alpha }\left\vert \Psi \right\rangle =d_{\alpha }\left\vert \Psi
\right\rangle =0,\qquad \left[ c_{\alpha },\overline{c}_{\beta }\right] =%
\left[ d_{\alpha },\overline{d}_{\beta }\right] =\delta _{\alpha \beta }
\end{equation}%
with
\begin{equation}
\begin{split}
a_{\alpha }& =\sqrt{L}A_{\alpha }+c_{\alpha },\qquad a_{\alpha }^{\dagger }=%
\sqrt{L}\overline{A}_{\alpha }+\overline{c}_{\alpha } \\
b_{\alpha }& =\sqrt{L}B_{\alpha }+d_{\alpha },\qquad b_{\alpha }^{\dagger }=%
\sqrt{L}\overline{B}_{\alpha }+\overline{d}_{\alpha }.
\end{split}%
\end{equation}
Note that the new hatted creation operators are not the hermitian conjugate
of the annihilation operators due to the non-hermiticity of the Liouvillian.

Substituting the shifted operators into the Liouvillian\ %
\eqref{eq:liouv_schwinger_rep} and expanding it in orders of\ $L$,
\begin{equation}
\mathcal{L}=L\mathcal{L}_{1}+\sqrt{L}\mathcal{L}_{1/2}+\mathcal{L}_{0}+%
\mathcal{O}\left( L^{-1/2}\right),
\end{equation}%
it is straightforward to see that the choice \ $A_{\alpha }=\overline{A}%
_{\alpha }=B_{\alpha }=\overline{B}_{\alpha }=\delta _{\alpha ,\eta }$ and\ $\mu_a = \gamma - i\varepsilon_{\eta},\ \mu_b = \gamma + i\varepsilon_{\eta}$, for
an arbitrary value of \ $\eta =1,\cdots, N$, brings the first two terms in the
Liouvillian $\mathcal{L}_{1}$ and\ $\mathcal{L}_{1/2}$ to zero. Moreover,
this solution also satisfies
\begin{equation}
s_{\alpha }=\langle \overline{\Psi }|a_{\alpha }^{\dagger }a_{\alpha
}-b_{\alpha }^{\dagger }b_{\alpha }|\Psi \rangle =0
\end{equation}%
which corresponds to the sector of the SS. We would naively
conclude that the SS is $N$ times degenerate. However, we will see in the next order correction ($\mathcal{L}_{0}$) that only one of these solutions defines a SS with 0 eigenvalue.

\begin{figure}[t]
	\includegraphics[width=.66\textwidth]{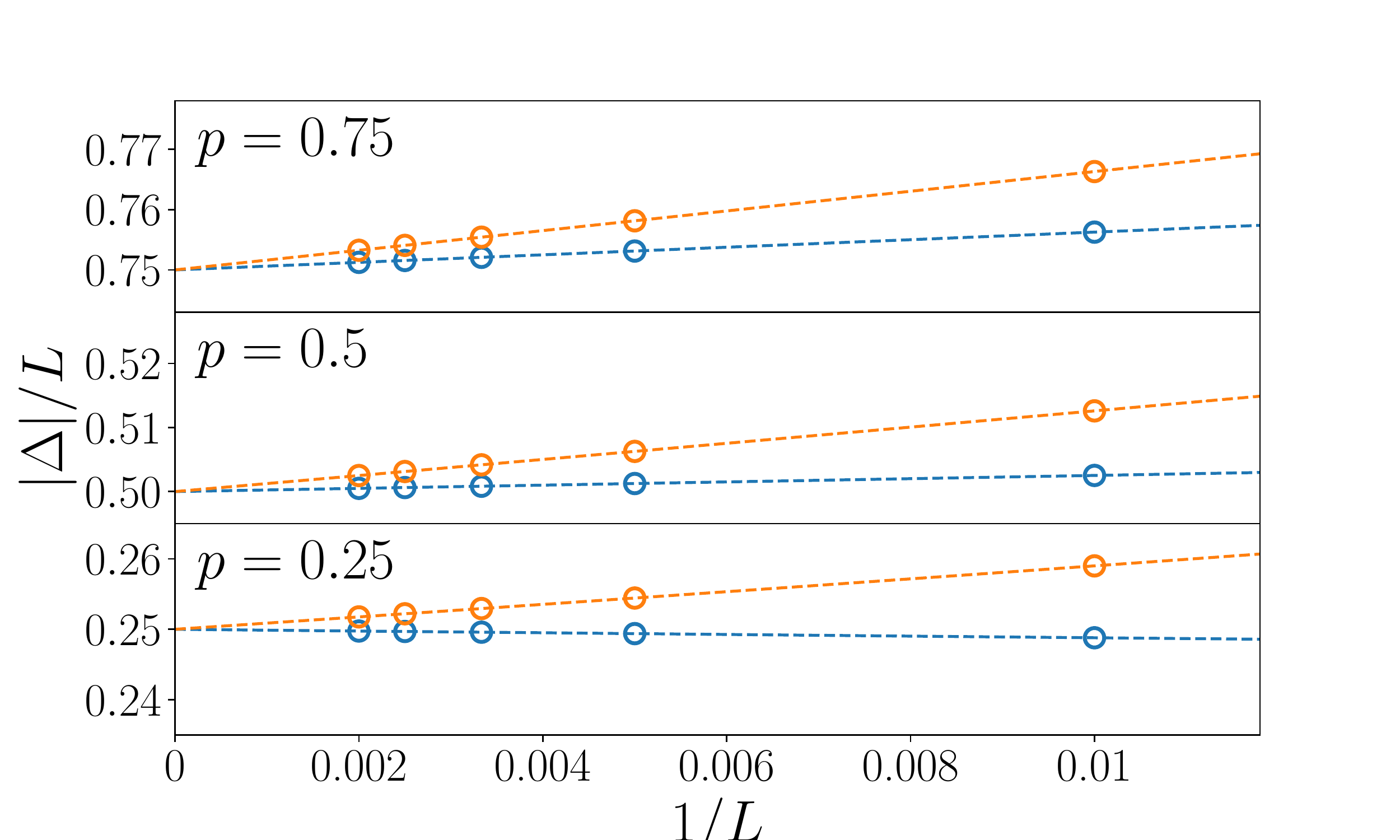}
	\caption{Finite size interpolation of the dissipative gap of the $SU(3)$ Liouvillian. From top to bottom, we show the rescaled real part of the closest to zero eigenvalue  of the symmetry sectors\ $s=(1,-1,0)$ (blue) and\ $s=(1,0,-1)$ (orange) for different Liouvillian sizes\ $L$ and values of\ $p$. We  perform a polynomial fit up to fourth order in\ $1/L$ of these results (dashed lines) and extrapolate the lowest Liouvillian eigenvalues to the limit\ $L\rightarrow \infty$ . The fit shows that the dissipative gap is degenerate in this limit, with a value of\ $|\Delta|/L = p\Gamma$. Data shown for\ $\Gamma=\Gamma_{0}=1$, and $L=100, 200, 300, 400, 500$.}
	\label{fig:dissipative_gap}
\end{figure}

With that choice of the shift parameters we can write the next order of
the Liouvillian $\mathcal{L}_{0}$ as
\begin{equation}
\begin{split}
\mathcal{L}_{0}=& \sum_{\alpha \left( \neq \eta \right) }\left( i\varepsilon
_{\alpha } - i\varepsilon_{\eta} - \gamma \right) \overline{d}_{\alpha
}d_{\alpha }-\sum_{\alpha \left( \neq \eta \right) }\left( i\varepsilon
_{\alpha }- i\varepsilon_{\eta} + \gamma \right) \overline{c}_{\alpha
}c_{\alpha }-\gamma \left( N-1\right) +\gamma p\left( N+1-2\eta \right)  \\
& +\gamma \left( 1-p\right) \sum_{\alpha \left( >\eta \right) }\overline{c}%
_{\alpha }\overline{d}_{\alpha }+\gamma \left( 1+p\right) \sum_{\alpha
	\left( <\mu \right) }\overline{c}_{\alpha }\overline{d}_{\alpha }+\gamma
\left( 1-p\right) \sum_{\alpha \left( <\eta \right) }c_{\alpha }d_{\alpha
}+\gamma \left( 1+p\right) \sum_{\alpha \left( >\eta \right) }c_{\alpha
}d_{\alpha }.
\end{split}%
\end{equation}
The Liouvillian $\mathcal{L}_{0}$ is quadratic in the boson operators. Therefore, it can diagonalized by means of a non-unitary Bogoliubov
transformation that defines quasiboson excitations and provides the next order
correction to the SS (quasiboson vacuum)
\begin{equation}
\begin{split}
e_{\alpha }=& u_{\alpha }c_{\alpha }-v_{\alpha }\overline{d}_{\alpha },\quad
f_{\alpha }=u_{\alpha }d_{\alpha }-v_{\alpha }\overline{c}_{\alpha } \\
\overline{e}_{\alpha }=& \overline{u}_{\alpha }\overline{c}_{\alpha }-%
\overline{v}_{\alpha }d_{\alpha },\quad \overline{f}_{\alpha }=\overline{u}%
_{\alpha }\overline{d}_{\alpha }-\overline{v}_{\alpha }c_{\alpha },
\end{split}%
\end{equation}%
with\ $u_{\alpha }\overline{u}_{\alpha }-v_{\alpha }\overline{v}_{\alpha }=1$
so that\ $e_{\alpha },f_{\alpha },\overline{e}_{\alpha },\overline{f}%
_{\alpha }$ are canonical operators. This transformation puts\ $\mathcal{L}_{0}$
into a diagonal form with vacuum energy\ $C_\eta$
\begin{equation}
\mathcal{L}_{0}=C_\eta + \sum_{\alpha \left( \neq \eta \right) }\overline{e}_{\alpha
}e_{\alpha }\Big[ -i\left( \varepsilon _{\alpha }-\varepsilon _{\eta
}\right) -|p|\gamma \Big] +\overline{f}_{\alpha }f_{\alpha }\Big[ i\left(
\varepsilon _{\alpha }-\varepsilon _{\eta }\right) -|p|\gamma \Big].
\end{equation}%
The choice  $\eta =1$ for\ $p>0$ and $\eta =N$ for  $p<0$ makes\ $C_\eta = 0$, hence there is a unique boson coherent state (SS) that condenses in the level\ $\eta=1$ or $\eta =N$.

The dissipative gap\ $\Delta$ is given by\ $-|p|\gamma$ in the TL. Since the gap does not depend on  $\alpha$, there are\ $2\left(N-1 \right)$ states whose eigenvalues have the  same real  part, generating bands like those appearing in figure  \ref{fig:liouvillian_spectra}(a). The breaking of these bands  is a finite size effect which can be evaluated in the next order of $L$. We numerically prove this behavior for the $SU(3)$ case in figure ~\ref{fig:dissipative_gap}, where we perform a polynomial fit of fourth order to  the dissipative gap as a function of $1/L$ for two  symmetry sectors\ $s=(s_1,s_2,s_3)$. For $\Gamma=\Gamma_0=1$, we obtain the following fit parameters

\begin{equation}
\begin{split}
s=(1,-1,0):\quad\quad\quad \Delta/L &= -p\,\Gamma + \frac{\Gamma}{L}\left( \frac{1}{2} - \frac{3p}{2}\right) + \mathcal{O}\left(\frac{1}{L^2} \right)  \\
s=(1,0,-1):\quad\quad\quad  \Delta/L &= -p\,\Gamma + \frac{\Gamma}{L}\left( -\frac{1}{2} - \frac{3p}{2}\right) + \mathcal{O}\left(\frac{1}{L^2} \right),
\end{split}
\label{extra}
\end{equation}
with  maximum standard deviations of\ $10^{-8}$ for the zeroth order coefficient  and\ $10^{-5}$ for that of  order\ $1/L$. We confirm that both states have eigenvalues with the same real part in the TL ($-|p| \gamma$). It is at the next order in $1/L$ where the degeneracy  is broken, so that  the symmetry sector\ $s=(1,-1,0)$ contains the eigenvalue with  least negative real part, defining  the dissipative gap for finite Liouvillian sizes. As an example, for $L=10$ and $p=0.5$, the inset of panel b in figure  \ref{fig:liouvillian_spectra} and panels b and c of figure\ \ref{fig:exact_solution}(b,c) these two states with real part -5.297 [$s=(1,-1,0)$] and -6.388 [$s=(1,0,-1)$]. The corresponding extrapolated values from (\ref{extra}) are -5.5 and -6.25 respectively.

\section{Conclusions}
The most important characteristic and attribute of the RG models of any rank $r$ is the existence of a complete set of $M$ integrals of motion that contain $M+r$ free complex parameters. Any linear combination of the integrals of motion, adding a set of $M$ new free parameters, produces an exactly solvable many-body operator.
We have exploited this property to obtain exactly solvable Liouvillians of dissipative quantum systems. Making use, for the first time, of the $SU(N)$ trigonometric family of RG models of arbitrary rank, we could derive exactly solvable Liouvillians describing the
dynamics of open quantum multi-level atom systems. As the simplest example, we
showed that the dissipative collective spin model, whose exact solution was
presented in\ \cite{Rib-Pro}, can be derived from the $SU(2)$
trigonometric RG model. We then moved  to the next degree of complexity and  studied in depth  the rank 2 $SU(3)$ RG model that describes dissipative systems of 3-level atoms. We worked out in
detail the particular combination of trigonometric $SU(3)$ integrals of
motion that give rise to the desired Liouvillian, as well as  the corresponding
mathematical form of the eigenvalues depending on two sets of
spectral parameters that are fixed by the solutions of non-linear coupled RG equations.

Solving the  trigonometric $SU(3)$-RG equations for the full Liouvillian spectrum is more
challenging than in the $SU(2)$ case, where only one set of spectral parameters
appears and for which efficient numerical methods have been developed \cite{Claeys15}. In spite of the complexity, we were able to solve exactly and to study the properties of the solutions for the SS and several slow
decaying states for systems with a large number of atoms and various
Liouvillian parameters. These exact solutions were interpreted in terms of a two-dimensional
electrostatic-like equilibrium problem for the position of the spectral parameters in the complex plane.

We then studied the TL of the $SU(N)$ model making use of the Schwinger boson mean-field that we extended to treat non-hermitian operators. We checked that
the SS energy is 0 to first order in the inverse of the number of atoms $1/L$ and obtained the eigenvalues closest to the SS described by one quasiboson excitations.
 A scaling analysis from the
exact solution for finite but very large systems of three-level atoms confirmed the analytical
expressions of the mean-field Liouvillian eigenvalues. Moreover, this finite size study provided the next to leading  order correction to the closest eigenvalues,  which proved to be accurate even for small systems.

We hope that these new exactly solvable models of dissipative $N$-level atom systems will add up to the comprehension of the dynamics of open quantum systems. We also expect that the techniques used in this work could be applied to other RG models based on different semi-simple algebras to extend the realm of exactly solvable dissipative systems.

\bigskip

\textit{Acknowledgments---} J.D. and A.R. acknowledges financial support from the
Spanish Ministerio de Ciencia, Innovaci\'{o}n y Universidades and the
European regional development fund (FEDER), Project No.
PGC2018-094180-B-I00. S.L.-H. acknowledges financial support from Mexican
CONACyT project CB2015-01/255702. This collaboration has been supported by the
Spanish Grant I-COOP2017 Ref:COOPB20289.

\appendix

\section{Exact $SU(N)$ and $SU(3)$ wave function}
\label{app:wavefun}

In \cite{Ushveridze94}, the common  unnormalized eigenfunctions of the rational RG integrals are provided for arbitrary semi-simple Lie algebras. By extending that result to the trigonometric case and considering a particular RG  model of $M$ $SU(N)$ copies, $K_{\alpha\beta m}$, we obtain
$$
|\Psi\rangle=\prod_{i=1}^{M_1}\mathcal{K}_{21}\left(E_{i}^{(1)}\right)\prod_{j=1}^{M_2}\left[\mathcal{K}_{32}\left(E_{j}^{(2)}\right)+ \overleftarrow{\mathcal{I}_{2}^1}\left(E_{j}^{(2)}\right) \right]\times...\times \prod_{k=1}^{M_{N-1}}\left[\mathcal{K}_{N N-1}\left(E_{k}^{(N-1)}\right)+ \sum_{b=1}^{N-2}\overleftarrow{\mathcal{I}}_{N-1}^{b}\left(E_{k}^{(N-1)}\right) \right]|\Lambda\rangle
$$
where $$\mathcal{K}_{\alpha\beta}(x)=\sum_{m=1}^M X^*(z_m-x) K_{\alpha\beta,m}\qquad {\hbox{with}}\qquad   X^*(u)=\frac{e^{-\mathrm{i}u}}{\sin u},$$ and  $
|\Lambda\rangle=\bigotimes_{m=1}^M|\Lambda_m\rangle $ is the tensor product of  HW states.
The operators
\begin{equation*}
\overleftarrow{\mathcal{I}}_{a}^{b}\left(E_j^{(a)}\right)=\sum_{i'=1}^{M_b}X^*\left(E_{i'}^{(b)}-E_{j}^{(a)} \right)
\overleftarrow{I}_{a i'}^b
\end{equation*}
act upon the space of operators $\mathcal{K}_{\alpha\beta}\left(E_j^{(a)}\right)$, through  $\overleftarrow{I}_{a i'}^b$ defined as
$$
\mathcal{K}_{\alpha \beta}\left(E_i^{(\beta)}\right)\overleftarrow{I}_{a i'}^b=\delta_{\alpha a}\delta_{\beta b}\delta_{ii'}\mathcal{K}_{\alpha+1 \beta}\left(E_i^{(\beta)}\right).
$$
Observe that the  wave function is completely determined by the sets of spectral parameters $E_i^{(a)}$.

For the particular case of the Liouvillian of three-level atoms, which consists of  two $SU(3)$ copies with HW states $|\Lambda\rangle=|L00\rangle\otimes|00L\rangle$ and $(z_1,z_2)=(0,z)$, the unnormalized eigenfunctions simplify to
\begin{equation}
\prod_{i=1}^{M_1}\bar{\mathcal{K}}_{21}(e_i)\prod_{j=1}^{M_2}\left[\frac{\mathrm{i}(p^{-1}-1)%
}{\omega_j-\mathrm{i}p^{-1}}K_{32,2}+\sum_{i^{\prime}=1}^{M_1}\frac{%
e_{i^{\prime}}-\mathrm{i}}{\omega_j-e_{i^{\prime}}}\overleftarrow{\mathcal{I}%
}_{2i^{\prime}}^1 \right] |\Lambda\rangle,
\end{equation}
where we have used variables $e_i$ and $\omega_j$, the fact that $K_{32,1}|\Lambda\rangle=0$, and we have redefined operators $$
\bar{\mathcal{K}}_{21}(e_i)=\left( K_{21,1}+\frac{\mathrm{i}(p^{-1}-1)}{e_i-\mathrm{i}%
p^{-1}}K_{21,2}\right)
$$ by discarding a   multiplicative factor.  Notice  that  contrary to the operator $K_{32,1}$ of the first copy and  due to the transformation $K_{\alpha\beta,2}=J_{\alpha\beta}=-\bar{J}_{\beta\alpha}$, the  operator of the
second copy $K_{32,2}$ does not annihilate the HW state, $K_{32,2}|\Lambda\rangle\not=0$.

\section{Exact diagonalization of the $SU(3)$ Liouvillian}
\label{app:exact_diagonalization}

Let us start with one copy of the $su(3)$ algebra. Within the \textit{irrep} with $C^1=L$ the complete set of states can be generated acting with the rising operators onto the HW state
\begin{equation}
|L : k_{2}, k_{3}\rangle \equiv \frac{1}{\sqrt{\mathcal{N}_{k_{2},k_{3}}}}
K^{k_{2}}_{21} K^{k_{3}}_{31} |L,0,0\rangle,
\end{equation}
with the restriction\ $0 \leq k_{2}+k_{3} \leq L$. We use the Schwinger representation (\ref{Sch1}) to compute the norm of this state
\begin{equation}
\begin{split}
|L:k_{2},k_{3}\rangle &= \frac{1}{\sqrt{\mathcal{N}_{k_{2},k_{3}}}}
\left(a^\dagger_2\right)^{k_{2}} \left(a^\dagger_3\right)^{k_{3}}
a_{1}^{-k_{2}-k_{3}} \frac{1}{\sqrt{L!}}\left(a^\dagger_1\right)^{L} |0\rangle
\\
&= \frac{1}{\sqrt{\mathcal{N}_{k_{2},k_{3}}}} \frac{\sqrt{L!}}{%
L-k_{2}-k_{3}!}\left(a^\dagger_2\right)^{k_{2}}
\left(a^\dagger_3\right)^{k_{3}} \left(a^\dagger_1\right)^{L-k_{2}-k_{3}}
|0\rangle,
\end{split}%
\end{equation}
therefore
\begin{equation}
\mathcal{N}_{k_{2},k_{3}} = L!\frac{k_{2}!k_{3}!}{k_{1}!},
\end{equation}
with\ $k_{1} = L - k_{2} - k_{3}$. Because the $SU(3)$ Liouvillian (\ref%
{Liuv}) contains two copies of the $su(3)$ algebra, we define the double basis
of states
\begin{equation}
|L: k_2, k_3; \overline{j}_2, \overline{j}_3\rangle \equiv \frac{1}{\sqrt{\mathcal{N}_{k_2, k_3}\,
\mathcal{N}_{j_{2}, j_{3}}}} K^{k_{2}}_{21}\, K^{k_{3}}_{31}\, \overline{J}^{j_{2}}_{21}
\,\overline{J}^{j_{3}}_{31}|L,0,0; L,0,0\rangle,
\end{equation}
with dimension\ $\left[ \left(L+1\right)\left(L+2\right)/2\right]^{2}$.

The operators \ $S_{\alpha}$\ \eqref{Cs} are weak symmetries of the Liouvillian, $[S_{\alpha},\mathcal{L}]=0$, defining the quantum numbers \ $s_{\alpha} = k_{\alpha} - \overline{j}_{\alpha}$. Therefore, we classify the  basis in blocks with quantum numbers\ $s_{2}, s_{3}$ (\ $s_{1}$ is is fixed by the constraint \ $\sum_{\alpha } s_{\alpha}=0$)
\begin{equation}
|L, s_2, s_3: k_2, k_3\rangle \equiv |L: k_2, k_3; k_2-s_2, k_3-s_3\rangle,
\end{equation}
whose dimension is \ $\left(L-s_{max}+1
\right)\left(L-s_{max}+2 \right)/2$, with\ $s_{max} = \max \left\lbrace
|s_2|, |s_3|, |s_2+s_3|\right\rbrace$.

The matrix elements of the Liouvillian in this basis are (we simplify the notation \ $|k_2, k_3\rangle \equiv |L,
s_2, s_3: k_2, k_3\rangle$)
\begin{equation}
\langle k_2, k_3|\mathcal{L}|k_2, k_3\rangle = -\Gamma C^2 - i\sum_\alpha
\varepsilon_\alpha s_\alpha - \frac{\Gamma_{0}}{2}\sum_\alpha s^2_\alpha -
\Gamma p \left(2k_3 - 2k_1 - s_3 + s_1\right)
\end{equation}
\begin{equation}
\begin{split}
\langle k_2-1, k_3|\mathcal{L}|k_2, k_3\rangle &= -\Gamma\sqrt{1+p} \sqrt{%
\left(k_1+1\right)k_2\left(s_1-k_1+1\right)\left(s_2-k_2\right)} \\
\langle k_2, k_3-1|\mathcal{L}|k_2, k_3\rangle &= -\Gamma\sqrt{1+p}\sqrt{%
\left(k_1+1\right)k_3\left(s_1-k_1+1\right)\left(s_3-k_3\right)} \\
\langle k_2+1, k_3-1|\mathcal{L}|k_2, k_3\rangle &= -\Gamma\sqrt{1+p}\sqrt{%
\left(k_2+1\right)k_3\left(s_2-k_2+1\right)\left(s_3-k_3\right)} \\
\langle k_2+1, k_3|\mathcal{L}|k_2, k_3\rangle &= -\Gamma\sqrt{1-p}\sqrt{%
k_1\left(k_2+1\right)\left(s_1-k_1\right)\left(s_2-k_2+1\right)} \\
\langle k_2, k_3+1|\mathcal{L}|k_2, k_3\rangle &= -\Gamma\sqrt{1-p}\sqrt{%
k_1\left(k_3+1\right)\left(s_1-k_1\right)\left(s_3-k_3+1\right)} \\
\langle k_2-1, k_3+1|\mathcal{L}|k_2, k_3\rangle &= -\Gamma\sqrt{1-p}\sqrt{%
k_2\left(k_3+1\right)\left(s_2-k_2\right)\left(s_3-k_3+1\right)} \\
\end{split}%
\end{equation}

\bibliographystyle{apsrev4-1}
\bibliography{references}

\end{document}